\documentstyle{cupconf}

\ifoldfss
\else
  \ifnfssone
    \newmathalphabet{\mathit}
      \addtoversion{normal}{\mathit}{cmr}{m}{it}
      \addtoversion{bold}{\mathit}{cmr}{bx}{it}
    \newmathalphabet{\mathcal}
      \addtoversion{normal}{\mathcal}{cmsy}{m}{n}
    \else
    \ifnfsstwo
    \fi
  \fi
\fi

\def\hexnumber#1{\ifcase#1 0\or1\or2\or3\or4\or5\or6\or7\or8\or9\or
 A\or B\or C\or D\or E\or F\fi }

\makeatletter
\ifx\CUP@mtlplain@loaded\undefined
\else
\fi
\makeatother
 \makeatletter
 \ifx\CUP@mtlplain@loaded\undefined
   \font\tenbmi=cmmib10 at 10pt
   \font\sevenbmi=cmmib10 at 7pt
   \font\fivebmi=cmmib10 at 5pt

   \newfam\bmifam
   \textfont\bmifam=\tenbmi
   \scriptfont\bmifam=\sevenbmi
   \scriptscriptfont\bmifam=\fivebmi
   
 \fi
 \makeatother
\ifnfsstwo

\fi
\ifnfssone

\fi
\ifoldfss

\fi

\mathchardef\varLambda="0103

\makeatletter
\ifx\CUP@mtlplain@loaded\undefined
\else
\fi
\makeatother
\def\etal{\mbox{\it et al.}}

\title{Spectral Tests of Models for Accretion Disks Around Black Holes}

\author[J.H. Krolik]
{J\ls U\ls L \ls I\ls A\ls N\ns H.\ns K\ls R\ls O\ls L\ls I\ls K\ls$^1$}

\affiliation{$^1$Department of Physics and Astronomy, Johns Hopkins University,
Baltimore MD 21218, USA}

\begin{document}
\ifnfssone
\else
  \ifnfsstwo
  \else
    \ifoldfss
      \let\mathcal\cal
      \let\mathrm\rm
      \let\mathsf\sf
    \fi
  \fi
\fi

\maketitle

\begin{abstract}
\end{abstract}

\firstsection

\section{Introduction: the Zeroth Order Picture}

    To test our ideas about the dynamics of accretion disks around
black holes, we compare the
radiation we receive from them with the radiative output predicted by our
models.  Standard models predict a very simple spectrum---a quasi-thermal
continuum.  However, even the most cursory glance at real accretion
disks immediately reveals that, while there often is a spectral component
resembling the expected quasi-thermal one, substantial energy is also
released in quite different ways---in hard X-rays from ``coronal" gas,
and also in extremely non-thermal radiation from relativistic
jets.  In this review I will not discuss how the energy is diverted into
coron\ae\ and more strongly non-thermal channels, or how well our models
for the radiation from these structures matches what is seen; the
{\it ad hoc} elements are so strong in these models that matching them
to observations does not provide strong tests of accretion disk dynamics.
Instead, I will concentrate on spectral properties of the quasi-thermal
continuum.  Here the character of the emergent radiation is directly
tied to the structure of the disk, and can in principle provide strong
diagnostics of how this structure is dynamically regulated.

    It is best to begin with the simplest picture: a quiescent, smooth,
geometrically thin disk that is in local thermodynamic equilibrium everywhere
from the marginally stable radius out to its outermost boundary.  Most of
the accretion energy is released at radii near $\sim 10 GM/c^2$ (the
radius of maximum dissipation is somewhat greater than the marginally
stable radius because relativistic effects and the outward transport
of energy associated with the angular momentum flux reduce the
depostion of heat in the innermost part of the disk).  The
characteristic temperature at which this energy is radiated is then
\begin{equation}
T_* \sim 2 \times 10^7 {\dot m}^{1/4} m^{-1/4} \hbox{K},
\end{equation}
where $\dot m$ is the accretion rate relative to the rate which would
produce an Eddington luminosity, and $m$ is the mass of the black hole
in solar units.  Thus, we expect stellar black holes to radiate primarily
soft X-rays, while galactic scale black holes (as in active galactic nuclei)
should radiate primarily in the ultra-violet.

    This description of the spectrum can be easily refined by integrating
over the actual distribution of disk dissipation.  Assuming LTE, the
temperature falls towards larger radius $r$ as $r^{-3/4}$.  The sum of
all the local black bodies then gives an overall luminosity per unit
frequency
\begin{equation}
L_\nu \propto \nu^{1/3} \exp\left(-h\nu/kT_*\right),
\end{equation}
where the characteristic temperature $T_*$ is approximated by the estimate
of equation 1.1 (a more careful calculation would be more punctilious about
the exact location of the maximum dissipation rate).

    The approximate expression given in equation 1.2 can be easily compared
with real spectra to gain a sense of perspective about how well our
expectations are vindicated in real objects.  Ironically, the quality of
data at our disposal is much better for AGN than for Galactic black holes.
This is because the band in which the thermal emission peaks for
Galactic black holes ($< 1$~keV) is one for which, at least hitherto,
spectroscopy has been relatively undeveloped, while in the ultraviolet,
where the thermal component of AGN emission peaks, good quality spectroscopy
is relatively easy to obtain.  As Figure 1 shows, while the UV portion of
the spectrum is, to zeroth order, replicated by the simple sum-of-blackbodies
spectrum, our simple models of accretion disks don't come close to predicting
the extremely broad-band emission seen in these objects.

\begin{figure}

\vspace{10cm}

\includegraphics{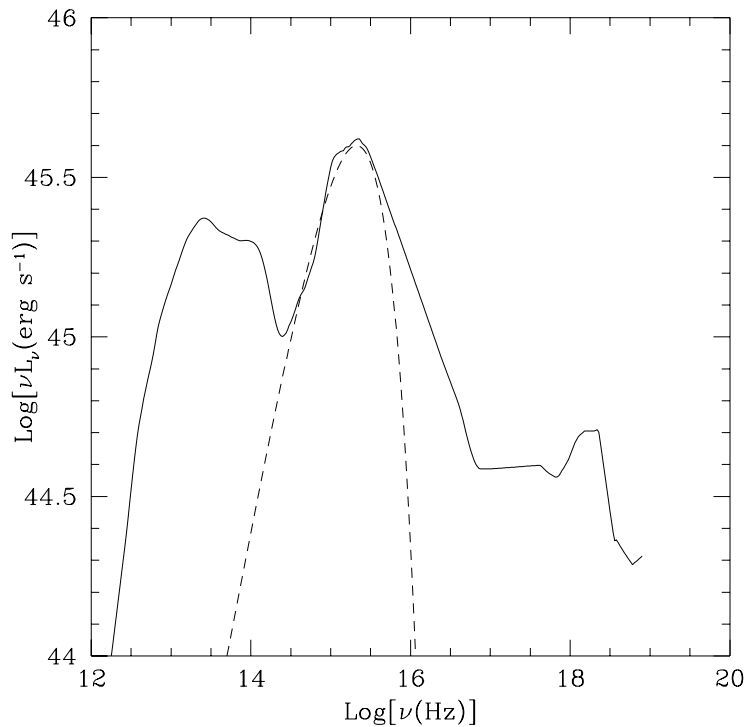}

\caption{This failure of standard accretion disk models is especially
prominent in AGN.  Here the composite spectrum of radio-quiet quasars
assembled by Elvis {\it et al.} (1994) is shown by the solid line, while
a spectrum with the shape of equation 1.2 for $T_* = 7 \times 10^4$~K is
shown by the dashed line.}

\end{figure}

\section{Issues in Detailed Modelling}

    Given this much success, the next question to ask is whether more
detailed predictions are equally successful.  However, to make these
predictions, we must first take a closer look at the accretion disk
model under consideration.  In particular, it contains at least two
glaring omissions: we have ignored all relativistic corrections, yet
by definition most of the light comes from a region whose gravitational
potential is relativistically deep; and we have no assurance that the
locally radiated spectrum is Planckian.

    First consider the latter question.  To answer it, we must solve three
separate problems: the equilibrium radial structure of the disk (to find
the optical depth as a function of radius); the equilibrium vertical
structure (to find the temperature and density as a function of altitude
at each radius); and the radiation transfer problem
(to find the local emergent spectrum as a function of viewing angle).
Unfortunately, there are major gaps in our understanding in each of
these three areas.  As a result, in each of them it is necessary
to make some unjustified assumption in order to proceed.  Consequently,
{\it every} detailed prediction of accretion disk spectra is model-dependent in
important ways.

\subsection{Radial structure}

    The fundamental equation governing the radial equilibrium is conservation
of angular momentum:
\begin{eqnarray}
- \int \, dz \, T_{r\phi} &= {\dot M \Omega \over 2\pi} R_T (r) \\
   &= {2 \mu_e c^2 \over e \sigma_T} {\dot m \over x^{3/2}} R_T,
\end{eqnarray}
where $T_{r\phi}$ is the $r-\phi$ component of the stress tensor, $\dot M$
is the mass accretion rate, $\Omega$ the local orbital frequency, and $R_T (r)$
is a correction factor which accounts for both relativistic effects and
the outward transport of angular momentum [\cite{NT73}].  The dimensionless
form given
in the second line is often the more useful one.  In it, $\mu_e$ is the mass per
electron, $e$ is the radiative efficiency of accretion in rest-mass units,
$\sigma_T$ is the Thomson cross section, and $x$ is the radius in gravitational
units ({\it i.e.}, the unit of length is $GM/c^2$).

      We do not yet know how to make a calculation from first principles for
$T_{r\phi}$ in terms of other disk quantities.  Instead, it has long been
popular to
appeal to dimensional analysis and guess that $T_{r\phi} = \alpha p$,
where $\alpha$ is a dimensionless constant and $p$ is the local pressure
(Shakura \& Sunyaev 1973).  Recent simulations of the nonlinear development
of magneto-rotational instabilities (Stone {\it et al.} 1996; Brandenburg
{\it et al.} 1996; Brandenburg, this volume) suggest that this may, in
fact, not be a bad approximation, and that $\alpha \sim 0.01$ -- 0.1, but
numerous uncertainties still remain.  For example, it is not entirely
clear {\it which} pressure sets the scale of the stress: gas pressure,
radiation pressure, or magnetic pressure?  Once this decision has been
made, an equilibrium may be found if one assumes steady-state accretion
and a smooth, geometrically thin structure for the disk.  Its surface
density is then $\propto \alpha^{-1}$, but it is also very sensitive to
which pressure sets the scale.   If the stress is proportional to
the total pressure, the optical depth of the disk can be much smaller
than if it scales with a pressure component that is only a fraction of
the total.  However, if $T_{r\phi}$ is proportional
to the total pressure, when $\dot m$ is more than a small fraction of unity,
these equilibria are both viscously (Lightman \&
Eardley 1974) and thermally (Shakura \& Sunyaev 1976) unstable.  Does this
mean that the ``$\alpha$-prescription" breaks down at the level of
perturbation theory?  Or does it mean that some other equilibrium should
be sought?

\subsection{Vertical structure}

   Equally troubling questions arise when considering the details of
vertical structure.  The amount of local dissipation per unit area in an
accretion disk is
\begin{eqnarray}
Q &= {3 \over 4\pi} {GM \dot M \over r^3} R_R (r) \\
  &= {3 \mu_e c^5 \over GM \sigma_T e} {\dot m \over x^3} R_R,
\end{eqnarray}
where $R_R$ is a correction factor that accounts for both
relativistic effects, and the outward transport
of energy associated with the angular momentum flux [also worked out
in \cite{NT73}].  However, we do not
know how this dissipation is distributed with height.  Clearly, the run of
temperature with altitude will be quite sensitive to the dissipation
distribution.

    In addition, when radiation pressure contributes significantly
to the vertical support of the disk against gravity (as commonly occurs),
there can be dramatic differences in structure depending on where the
radiation is created.  For example, if all the dissipation takes place in
the mid-plane, the radiation flux is constant with height, so that the
vertical radiation force is simply proportional to the local opacity.  On
the other hand, if the dissipation is confined to the disk surface, there is
no outward radiation flux in the body of the disk, and zero support against
gravity.  Moreover, if the dissipation takes place mostly within a few
optical depths of the surface, its exact distribution will clearly have
a major impact on the nature of the emergent radiation.

    When radiation pressure dominates the vertical support, even small
irregularities in the dissipation distribution can significantly affect
the outgoing spectrum.  Suppose, for example, that the dissipation in
the body of the disk is proportional to the gas density.  Then, because
the flux increases linearly with altitude in exact balance with the
increase of $g_z$ with altitude, the effective vertical gravity is nil,
and the density is constant as a function of height.  However, at the
very top of the disk this balance must be broken, for the density must
fall to zero.  The details of how sharply the density drops are very
sensitive to exactly where the dissipation occurs because the effective
gravity within the atmosphere depends on the balance between the
small additional gravity gained by a small rise in altitude and the
small additional radiation force due to the dissipation within that layer.
These details are important because the photosphere generally lies within
this region of the density roll-off.
    
     The issue of how the heat deposition varies from place to place
becomes still further confused when we consider the evidence that a
significant part of the total emisison comes out in hard X-rays, a portion
of which may then shine down on the disk.  In both AGN and Galactic black
hole systems, there are telltale signs of X-ray illumination in
the form of the ``Compton reflection bump":  Soft X-rays shining
on cool material suffer strong photoelectric absorption, so the albedo
of any accretion disk to photons from 0.5 -- 10~keV is generally quite
small, unless it is very thoroughly ionized.  On the other hand,
somewhat harder X-rays ($\sim 10$ -- 50~keV) are very readily reflected
by electron scattering because the highest energy photoionization edge
of any abundant element is that of Fe at 7.1~keV, and photoionization
cross sections drop rapidly with increasing energy above the edge.  The
reflection bump rolls over above 50~keV because higher energy photons
lose energy by Compton recoil as they are reflected.  These bumps
are often so prominent that the corollary absorption must contribute
significantly to disk heating.

    This external illumination
may substantially alter both the radial dependence of the integrated heating
rate and the vertical distribution of heating at a fixed radius.  Where
it is important, the heating is largely confined to a skin whose
thickness is at most a few Compton depths
({\it i.e.} $\sim 10^{24}$~cm$^{-2}$).  Precisely because the energy
is absorbed in such a thin layer, its effect there can be very strong.
Because the photosphere frequently lies at a comparable depth from
the surface, X-ray heating can have a major impact on features in the
emergent spectrum.

    In the limit that {\it most} of the heating occurs near the surface,
whether due to segregation of the internal dissipation or external
illumination, the instabilities endemic to radiation pressure-supported
disks are quenched (Svensson \& Zdziarksi 1994).   The reason is that
there is then little outgoing radiation flux within the body of the disk,
so it collapses to a state of substantially greater density (and also
greater total surface density).  Increases in the dissipation rate then
have no impact on the thickness of the disk, and the feedback which drives
the instabilities disappears.

\subsection{Radiation transfer}

     While the transfer problem is entirely understood in principle, in
practise it is so complicated that in every treatment so far, major
approximations have been made.  Thus, for technical, rather than
conceptual, reasons, the solutions are all, in one way or another,
model-dependent.  A major part of the art of constructing a good transfer
solution therefore lies in adroit choices of approximations.

     Certain features must surely be included in any transfer solution.  Thomson
opacity, for example, is almost always important.  Similarly, it is easy to
show that free-free opacity always affects at least the lower frequencies.
At AGN temperatures, HI and HeII photoionization opacity are certainly
significant, but at the higher temperatures of disks around stellar black holes,
H and He are equally surely fully stripped.

    Beyond this point, however, different
workers have made different choices, and some effects may be important in
certain ranges of parameter space, but not in others.  For example, when
computing the H and He photoionization opacity, it is necessary to make
some statement about the ionization balance for these elements.  The simplest
guess is that the ionization fractions are those given by the Saha equation,
but it is not obvious that this is always correct.  Indeed, one might expect
that the intensity of radiation in the H and HeII ionization continua would
{\it couple} to the fractional ionization of these species.  If one does
wish to compute the actual ionization balance of H and He, there are further
choices to be made about which processes are important and which are
negligible.  Ionizations can take place from excited states as well as
the ground state, so the excited state population balance must also be found.
How many states must be included?  And which processes?
In the most recent such calculation (Hubeny \& Hubeny 1997), the H atom
was permitted 9 different values of the principal quantum number, but all
states having the same principal quantum number were assumed to be in
detailed balance.  In addition (and probably more significantly),
only bound-free processes were considered.  This may have been a significant
oversight because at the densities prevalent in their disk model, the
bound-bound transition rates due to electron collisions can be comparable to
the bound-free rates.  Moreover, because non-LTE effects are very sensitive
to density, their character can depend very strongly on the model choices
made when calculating the vertical structure of the atmosphere.

   Another open question is the possible role of heavy element opacities.
In AGN accretion disks, where the temperature might be $\sim 10^5$~K and
the density $\sim 10^{14}$~cm$^{-3}$, the thermal equilibrium H neutral
fraction is extremely small, $\sim 10^{-8}$.  Partially-ionized
stages of the more abundant heavy elements are therefore much {\it more}
common than neutral H; the issue is whether the energy bands in
which they have substantial opacity are important to the energy flow.
Their ionization continua lie at relatively high energy (at least several
tens of eV); their resonance lines have rather lower energy, but the
importance of lines depends very strongly on the amplitude of turbulent motions
in the disk.  This last matter is, of course, central to the uncertainties
already discussed regarding shear stress.  In disks around stellar black
holes, the Saha equation predicts that the dominant ionization stages
will have ionization potentials 5 -- $10 kT$; however, there may be
enough representatives of species with ionization potentials factors of
a few smaller to significantly contribute to the opacity.  This issue, too,
also has implications for vertical structure when radiation pressure
is an important part of the disk's vertical support.

    A further question has to do with the possible effects of Comptonization.
Its importance is gauged by the parameter
\begin{equation}
y \equiv (4 kT/m_e c^2)\max(\tau_T,\tau_T^2),
\end{equation}
for Compton optical depth $\tau_T$.  Here the relevant Compton depth is only
that portion of the optical depth above
the effective photosphere.  When $y$ is at least
order unity, repeated Compton scatters may impart significant
additional energy to photons before they leave the atmosphere.
The magnitude of Comptonization is extremely sensitive to model choices
and parameters.  For
example, if the disk equilibrium is one in which the stress is proportional
to the total pressure, the temperature in the atmosphere is close to the
effective temperature, free-free absorption is the only absorptive opacity,
and the gas pressure scale height in the atmosphere
is comparable to the total disk thickness (an assumption likely to overestimate
the importance of Comptonization),
\begin{equation}
y \simeq 0.6 {\dot m}^{5/3} m^{-1/6} \left({x \over 10}\right)^{-5/2}
R_{R}^{3/2} R_{z}^{-2/3} {\omega^2 \over (1 - e^{-\omega})^{2/3}} .
\end{equation}
Here $\omega = h\nu/kT$, and $R_z$ is another relativistic correction
factor, this one adjusting the vertical component of the gravity (Abramowicz
\etal 1997).  The frequency-dependence of $y$ is a consequence
of the frequency-dependence of the effective photosphere's location.
Unfortunately, at the present state of the art, Comptonization can only
be treated within the diffusion equation for radiation transfer.  Consequently,
studies of Comptonization cannot also deal with questions involving
either the angular distribution of the emergent radiation or sharp spectral
features such as lines and edges.  The second limitation comes about
because these features are created by changes in the source function on
scales comparable to a scattering length; hence, the diffusion approximation
does not give an adequate description.

\subsection{Relativistic effects}

    A final element that must be included in any proper model (but which
is sometimes forgotten) is a proper accounting for relativistic effects.
In addition to the dynamical corrections (encapsulated in $R_R$, $R_T$,
and $R_z$), there are also relativistic photon propagation effects
[\cite{C75}].  Disk
material close to a black hole moves at speeds close to $c$; as seen
by a distant observer, its radiation is therefore Doppler boosted and beamed.
In addition, there are intrinsically general relativistic effects:
gravitational redshift, and photon trajectory bending.  The last effect
must be properly melded with the intrinsic angular radiation pattern
of disk material in its own rest frame.  Clearly, when (as will certainly
happen here), Doppler shifts of order unity occur, there can be a dramatic
impact on sharp spectral features such as lines and edges.

\section{Results to Date}

    There have been many efforts over the past fifteen years or more to
compute the spectra to be expected from accretion disks around black holes.
The methods used include:
\begin{itemize}

\item stitching together stellar atmosphere solutions (Kolykhalov \& Sunyaev
1984);

\item summing local blackbodies, but applying general relativistic photon
propagation effects (Sun \& Malkan 1989);

\item solving the frequency-dependent transfer problem, but assuming
that the atmosphere is scattering-dominated at all locations and at all
frequencies (Laor \& Netzer 1989);

\item solving the frequency-dependent transfer problem including non-LTE
effects in H and He, but with non-standard stress prescriptions
(St\"orzer \etal\ 1994, Hubeny \& Hubeny 1997);

\item solving the frequency-dependent transfer problem in LTE, but
studying the effects of various stress prescriptions (Sincell \& Krolik 1998)
or external illumination (Sincell \& Krolik 1997);

\item solving the diffusion/Kompaneets equations in order to study
Comptonization (Ross, Fabian \& Mineshige 1992; Shimura \& Takahara 1993;
D\"orrer \etal\ 1996).

\end{itemize}

    Because no one calculation is the most nearly complete, I will
try to display the range of possibilities by showing a variety of calculations,
each highlighting a different aspect of the problem.  To set the stage,
it's best to begin with how ``standard" models depend on parameters.
Figure 2 shows the scaling with central mass at fixed accretion rate
relative to Eddington according to a calculation which employed full transfer
solutions and general relativistic effects, but made the LTE approximation
for the H and He ionization balances, and ignored all heavier elements.
Although the luminosity increases in proportion
to mass at fixed $\dot m$, the radiating area increases faster,
$\propto m^2$.  That is to say, the temperature at the innermost ring
scales $\propto (\dot m/m)^{1/4}$.  Consequently, as this quantity falls,
the spectrum becomes softer and the Lyman edge goes steadily deeper and deeper
into absorption.   If, on the other hand, the central mass is held constant and
the accretion rate is varied (Figure 3), the temperature rises,
so increasing accretion rate both hardens the overall spectrum and throws
the Lyman edge from absorption into emission.  In the hottest disks, there
can be substantial flux in the HeII continuum, although the edge itself
generally stays in absorption for parameters appropriate to AGN.
  
\begin{figure}

\vspace{10cm}

\includegraphics{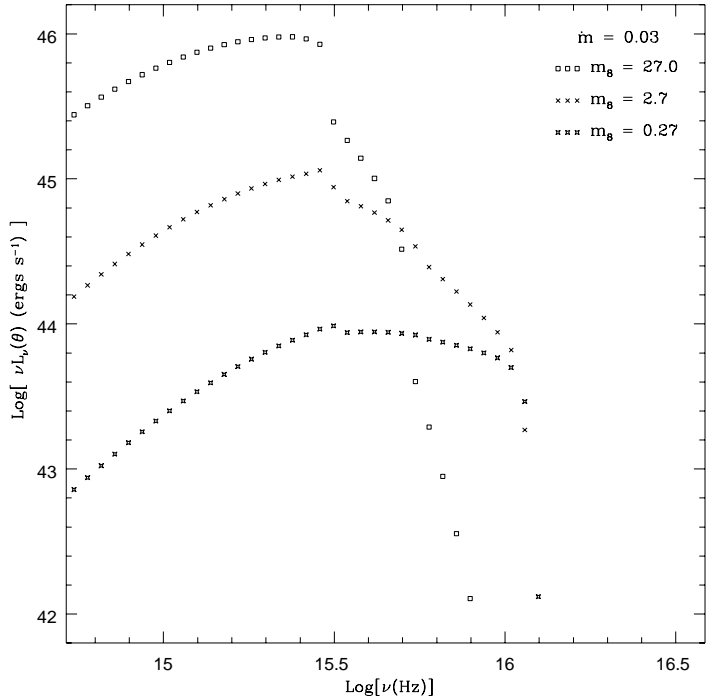}

\caption{The predicted spectra from accretion disks around
non-rotating supermassive black holes viewed pole-on.  In
all three cases shown, $\dot m = 0.03$, but the mass varies from
$m = 3 \times 10^7$ to $m = 3 \times 10^9$ (from Sincell \& Krolik 1998).}

\end{figure}
  
\begin{figure}

\vspace{10cm}

\includegraphics{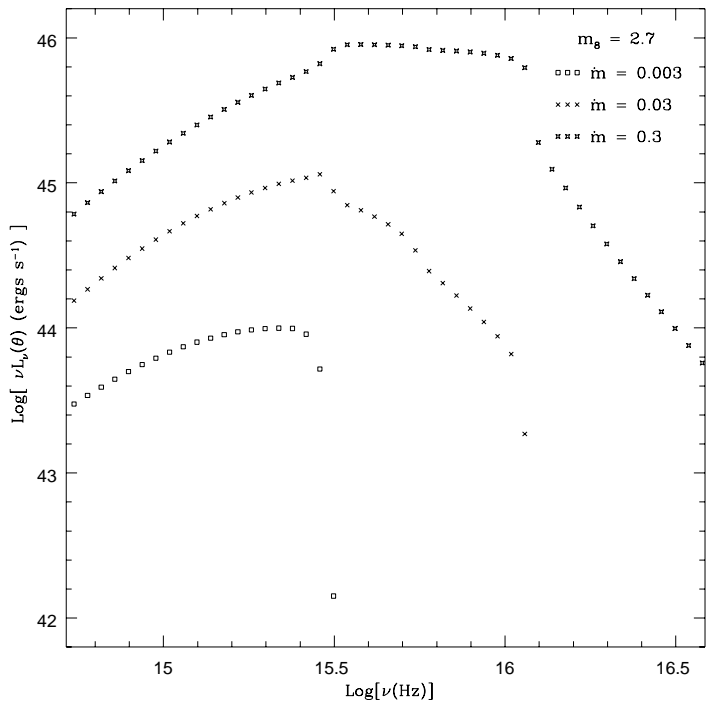}

\caption{The predicted spectra from accretion disks around
non-rotating supermassive black holes viewed pole-on.  In
all three cases shown, $m = 3 \times 10^8$, but the scaled
accretion rate varies from
$\dot m = 0.003$ to $\dot m = 0.3$ (from Sincell \& Krolik 1998).}

\end{figure}

\subsection{Dependence on radial structure model}

   With this background, we may now inquire into the effects of the
systematic uncertainties discussed in the previous section.  First, how serious are the effects of the uncertainty in the radial equilibrium?  Figure 4 shows
what happens when the stress prescription is changed
from $\alpha p_g$ to $\alpha \sqrt{p_g p_r}$ to $\alpha (p_r + p_g)$.
Greater stress at fixed accretion rate leads to both smaller surface density
and smaller volume density.  Here the accretion rate is high enough that
$p_r \gg  p_g$ in the interesting inner rings of the disk.  As a result,
the temperature in the atmosphere is lowest in the $\alpha p_g$ case and
highest when the stress is $\alpha (p_r + p_g)$.  At fixed total luminosity,
this leads to a generally harder overall spectrum for the disk with the
greatest stress, but the qualitative character of the spectrum changes
relatively little with different choices for the stress prescription.
  
\begin{figure}

\vspace{10cm}

\includegraphics{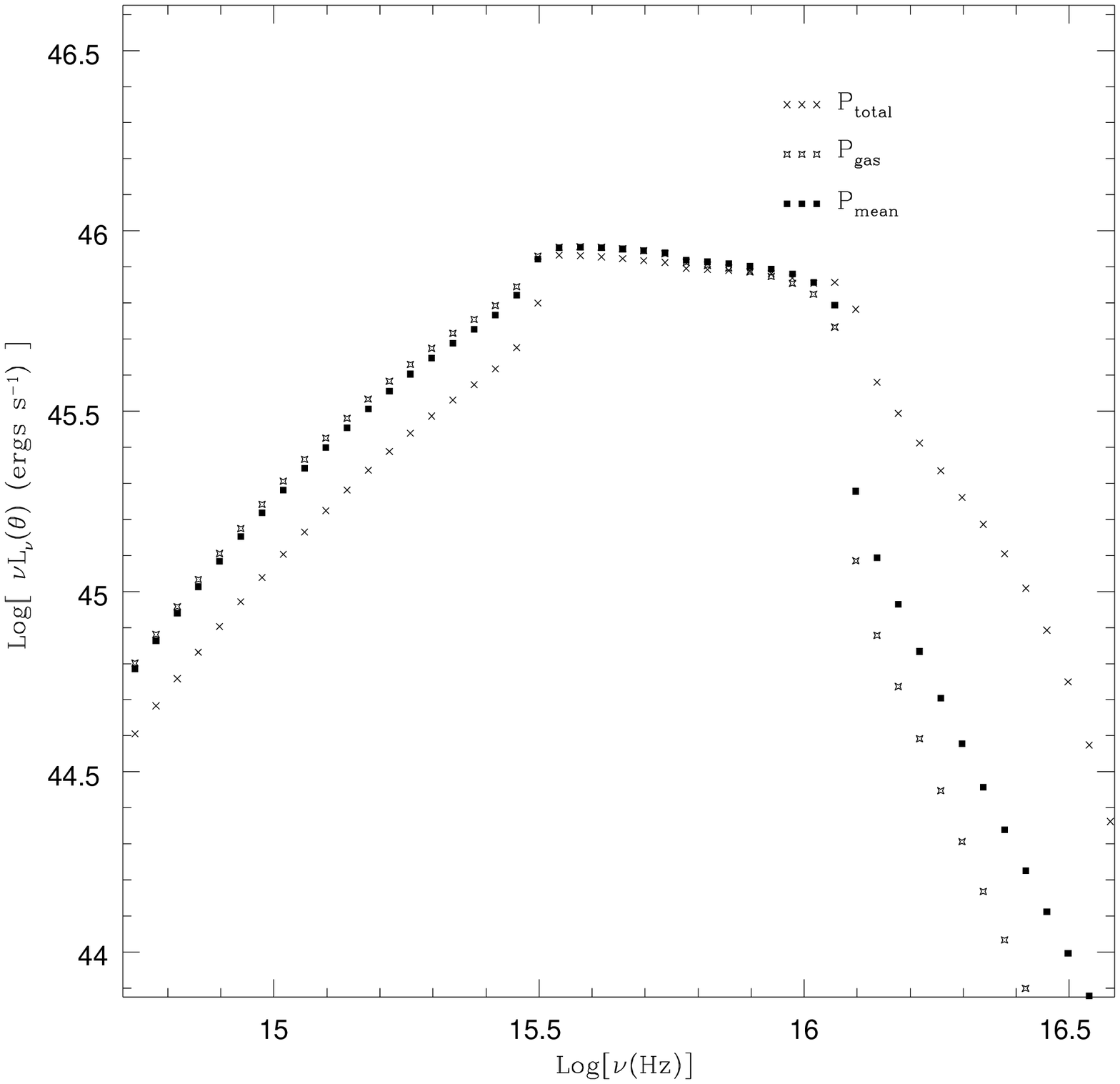}

\caption{The predicted spectrum from an accretion disk around a
non-rotating supermassive black hole viewed pole-on.  Here $\dot m = 0.3$
and $m = 3 \times 10^8$ (from Sincell \& Krolik 1998).}

\end{figure}

\subsection{Dependence on vertical distribution of heating}

    The next question to consider is the impact of the vertical distribution
of heat deposition.  In the most extreme limit, one in which all the
heating is concentrated into the disk atmosphere (as, for example,
X-ray heating may accomplish), the effects are dramatic.  Because
radiation pressure no longer supports the disk against the vertical
component of gravity $g_z$, it is geometrically much thinner.  Closer to
the disk midplane, $g_z$ is also smaller, so the gas pressure scale height
(at fixed temperature) is longer.  The density at the photosphere is then
smaller, even though the density deep inside the disk is much greater.  In
addition, because the heating rate per unit mass rises upward in
the X-ray-heated zone, so does
the temperature (although not necessarily monotonically).  The result is
emergent spectra that are {\it softer} than the corresponding spectra
from internally heated disks at low
frequencies, but with more flux at higher frequencies, and ionization
edges that are almost always in emission (compare Figure 5 to Figure 2).
  
\begin{figure}

\vspace{10cm}

\includegraphics{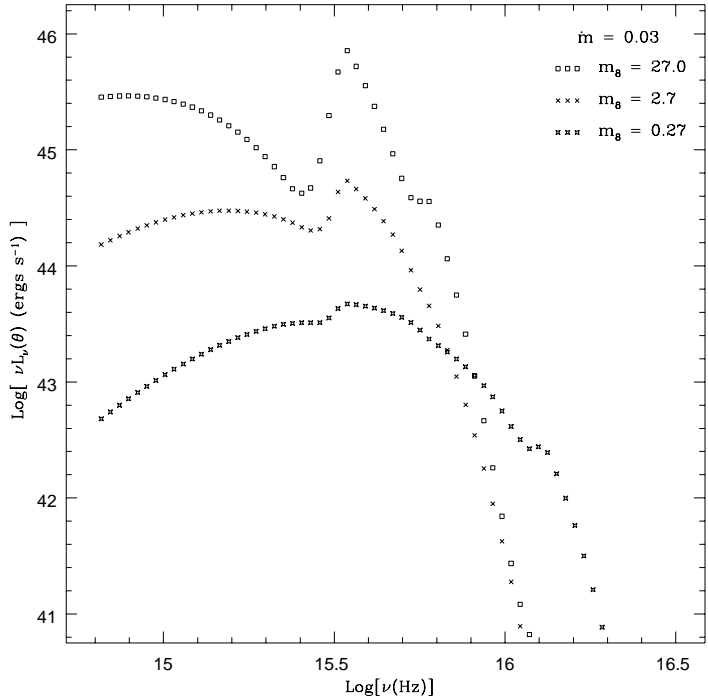}

\caption{The predicted spectra from X-ray-illuminated accretion disks around
non-rotating supermassive black holes viewed pole-on.   In
all three cases shown, $\dot m = 0.03$, but the mass varies from
$m = 3 \times 10^7$ to $m = 3 \times 10^9$ (from Sincell \& Krolik 1997).}

\end{figure}

\subsection{Implications of the LTE approximation}

    Different approximations in the radiation transfer solution are the
next issue whose effect we need to evaluate.  Is it necessary (or when
is it necessary) to compute the departure from LTE of the H and He
ionization balances?  \cite{S94} argued that when the photospheric
pressure was $ > 100$~dyne~cm$^{-2}$ ({\it i.e.}, the density
$n \sim 10^{13} T_{5}^{-1}$~cm$^{-3}$),
non-LTE effects are weak.  However, \cite{HH97} found that the LTE
approximation seriously misrepresented the strength of the HI and HeII
edges even when the photospheric pressure was several orders of magnitude
greater.  Part of the problem may be the definition of a non-LTE calculation:
\cite{HH97} included no bound-bound transitions in their evaluation of
the H and He excited state balances.  Another part of the problem (as
already discussed in \S 2.2) 
may be that the character of the non-LTE effects is exquisitely sensitive to
approximations in the solution of the hydrostatic equilibrium [an issue
which affects both \cite{S94} and \cite{HH97}].  Figure 6 illustrates
what the effects of non-LTE {\it might} be.  In this case,
both the radiative acceleration and $g_z$ were taken to be constant in
the atmosphere, 9 bound levels of HI, 8 of HeI, and 14 of HeII were
included, and all
bound-bound transitions were neglected.  Although the overall shape of
the continuum is not very sensitive to non-LTE effects,
they can completely change the nature of features at ionization edges. 
  
\begin{figure}

\vspace{10cm}

\includegraphics{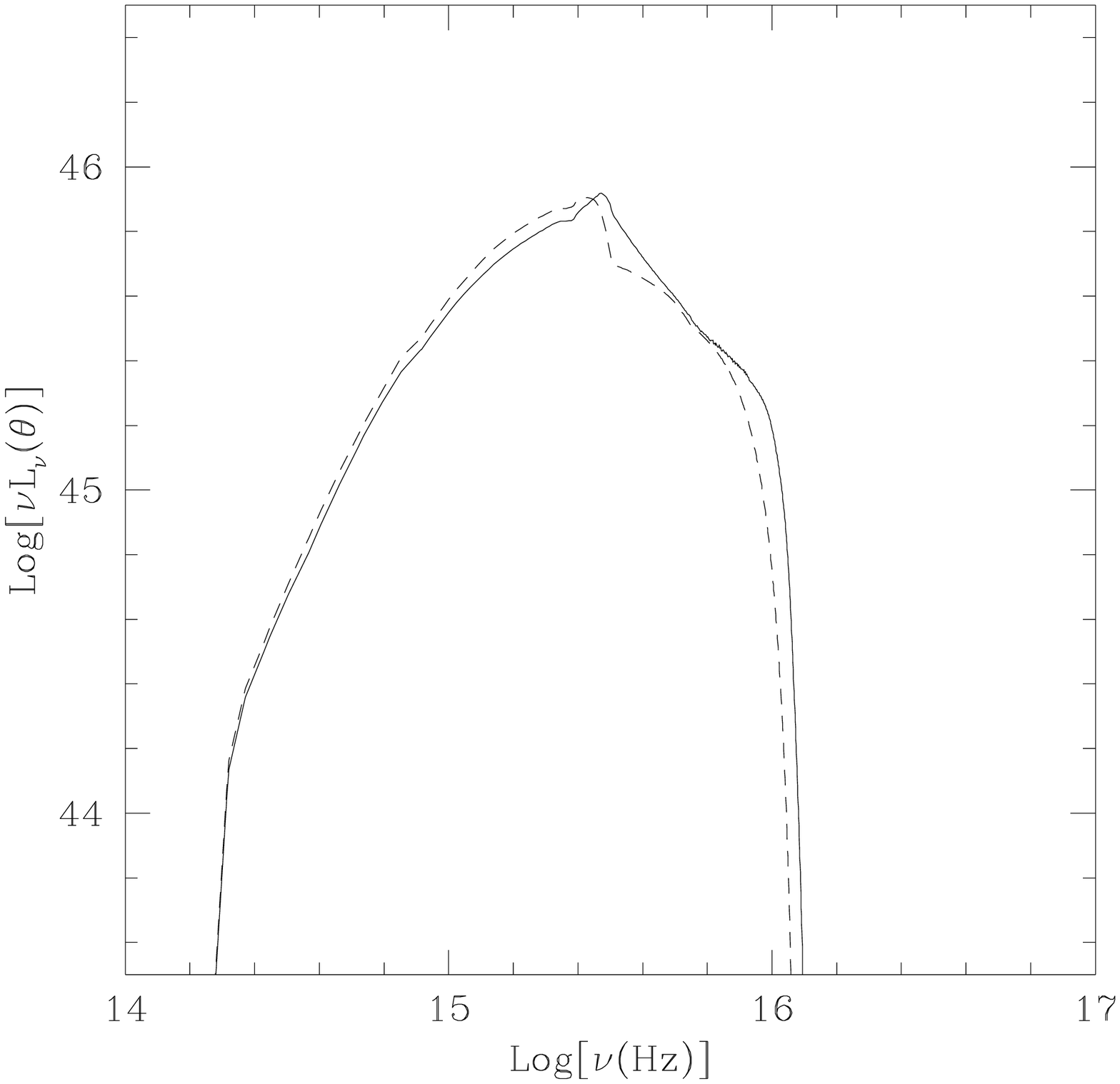}

\caption{LTE (dashed line) and non-LTE (solid line) spectra from
accretion disks around supermassive black holes viewed pole-on.   In these calculations (Agol 1997), $\dot m = 0.072$ and $m = 2 \times 10^9$,
and the normalized spin was 0.998.}

\end{figure}

\subsection{Influence of heavy element opacity}

     Particularly in AGN accretion disks, heavy elements may potentially
contribute substantially to the opacity, but these effects have just
begun to be explored.  To date there has been no calculation of how
this added opacity may affect the vertical structure of the disk, and
therefore $g_z$ in the atmosphere.  Indeed, because including all the
necessary atomic data is such a big job, and because the additional frequency
resolution required to follow the multitude of heavy element lines adds
such a large computational burden, there have been only partial
calculations of these effects even assuming a vertical structure computed
allowing only for H and He.  For this reason, the spectrum shown in Figure 7
[\cite{H98}]
should be regarded merely as a demonstration of the potential impact of
heavy element opacity.  It was computed on the basis of an atmosphere
solution whose vertical structure and temperature profile were determined
including only H and He.  Moreover, only those heavy element lines falling
between 600\AA\ and 1400\AA\ were used.  Nonetheless, the effects are large:
the integrated outgoing flux is reduced by roughly a factor of two.  An
atmosphere which includes heavy element opacities in a fully self-consistent
fashion will clearly look quite different from one with only H and He.

\begin{figure}

\vspace{10cm}

\includegraphics{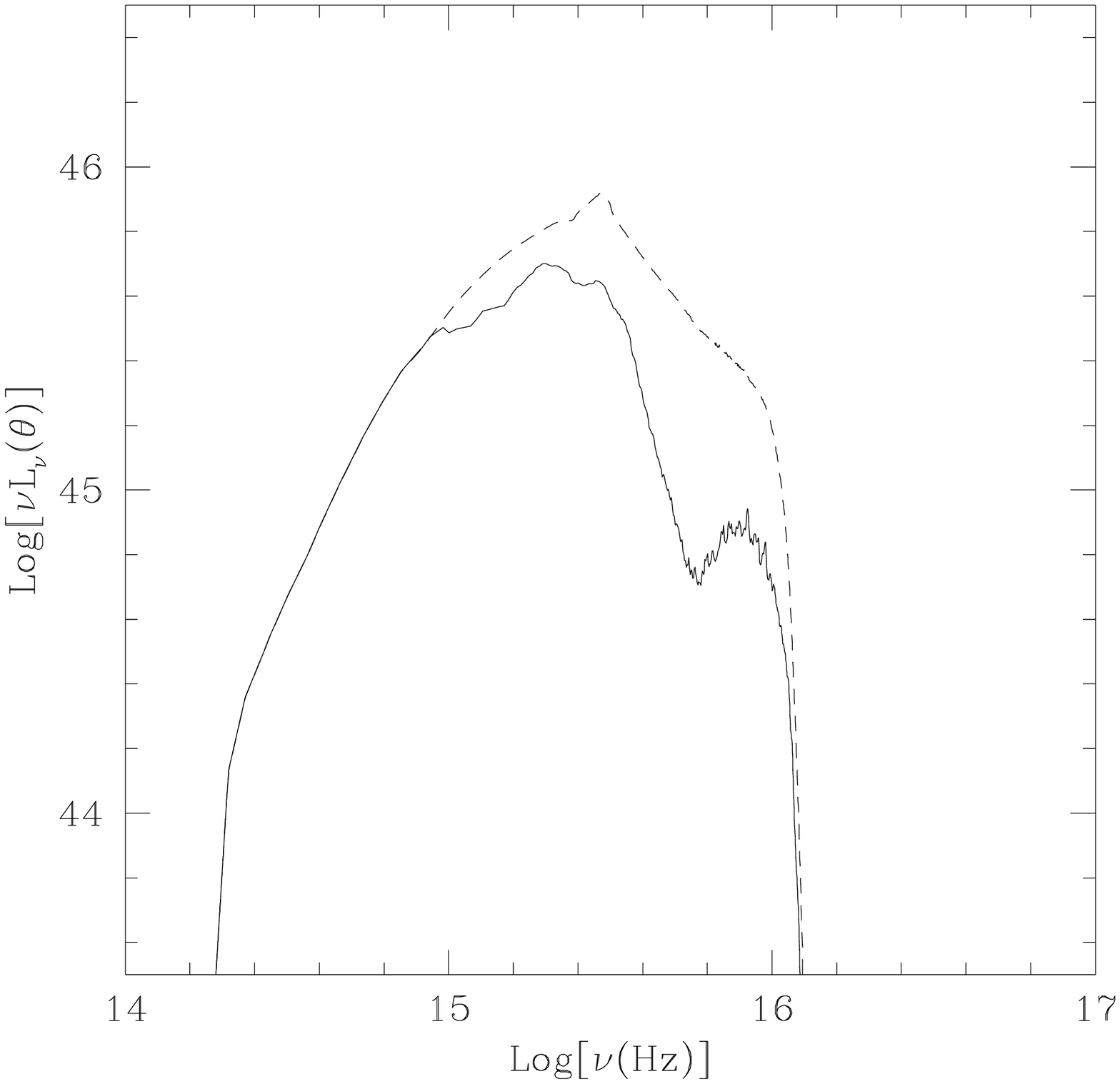}

\caption{Effects of heavy element opacity on the spectrum from an accretion
disk around a black hole with the same parameters as in Figure 6.
As described in the text, the
dashed line is a self-consistent disk spectrum whose opacity tables included
only H and He; the solid line was calculated using the disk structure and
temperature profile of the pure H and He model, but for predicting the
outgoing flux, metal features between 600\AA\ and 1400\AA\ were included.
The heavy element ionization fractions were assumed to be in LTE.}

\end{figure}

\subsection{Impact of relativistic effects}

       As discussed in \S 2.4, there are two separate classes of
relativistic effects: those entering the structural equations, and those
affecting the appearance of the disk to distant observers.  Both sorts
of corrections are relatively modest when the black hole has little spin
because the marginally stable orbit, at $x=6$, is relatively far from
the event horizon (at $x=2$).  However, as the spin increases, both
the marginally stable orbit and the event horizon move inward, both
approaching $x=1$ as the spin increases toward its maximum value.  Figure 8
shows how dramatic these effects can be.  All three predicted spectra
were calculated using the same physics (stress proportional to $\sqrt{p_r p_g}$,
LTE ionization fractions, no heavy elements) and pertain to the same $m$,
$\dot m$, and viewing angle.  When the central black hole spins rapidly,
sharp edge features (as could be seen in the predictions for the spinless
black hole) get stretched out over substantial frequency ranges.  The HeII
edge is stretched even farther than the HI edge because its origin is
confined more nearly to the innermost radii.  However, because the strongest
relativistic effect is the Doppler boosting and beaming of radiation into
the direction of orbital motion, these smoothing effects weaken considerably,
even for rapidly-spinning black holes, when the viewing angle is nearly
along the rotation axis.

\begin{figure}

\vspace{10cm}

\includegraphics{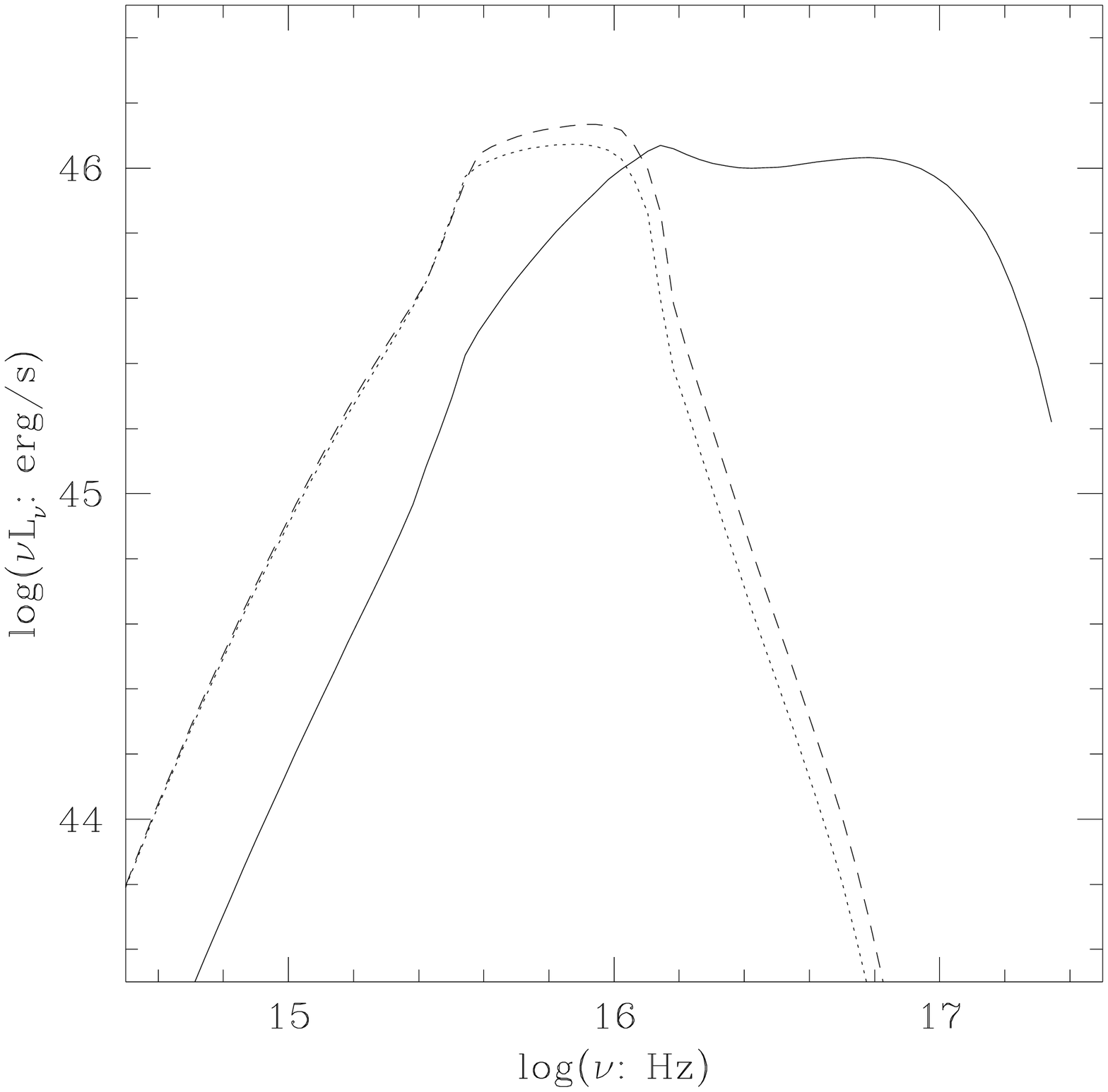}

\caption{Effects of general relativity on the observed spectrum.  All
three curves are computed for a disk with $\dot m = 0.3$ around a black
hole of mass $m = 3 \times 10^8$ viewed at an inclination angle of $40^\circ$.
However, the dotted curve is the prediction of a model in which the
black hole is spinless and the photons travel according to Newtonian rules,
while the dashed curve shows how the same disk looks in the real, {\it i.e.},
relativistic world, and the solid curve shows the spectrum from the same
disk if its central black hole had a normalized spin of 0.998.}

\end{figure}

\section{Comparison to observations}

    Now that we have seen the range of spectral shapes predicted by various
models, we can look at genuine observed spectra to see if any of their
features are predicted by the models.  Remarkably, the simplest
approximation---summing local blackbodies without regard for any of the
detailed physics we have discussed---does the best job of reproducing
the observations!  As shown in Figure 9, its slope comes very close
to matching the composite
slope (at least for the LBQS sample) at low frequencies, and, of course,
it is absolutely clean at all ionization edges, in excellent agreement
with the numerous observations showing that the Lyman edge is unobservably
weak in the great majority of quasars [\cite{K92}].  The zeroth-order
approximation still fails, however, to explain the strength of the
EUV continuum.

\begin{figure}

\vspace{10cm}

\includegraphics{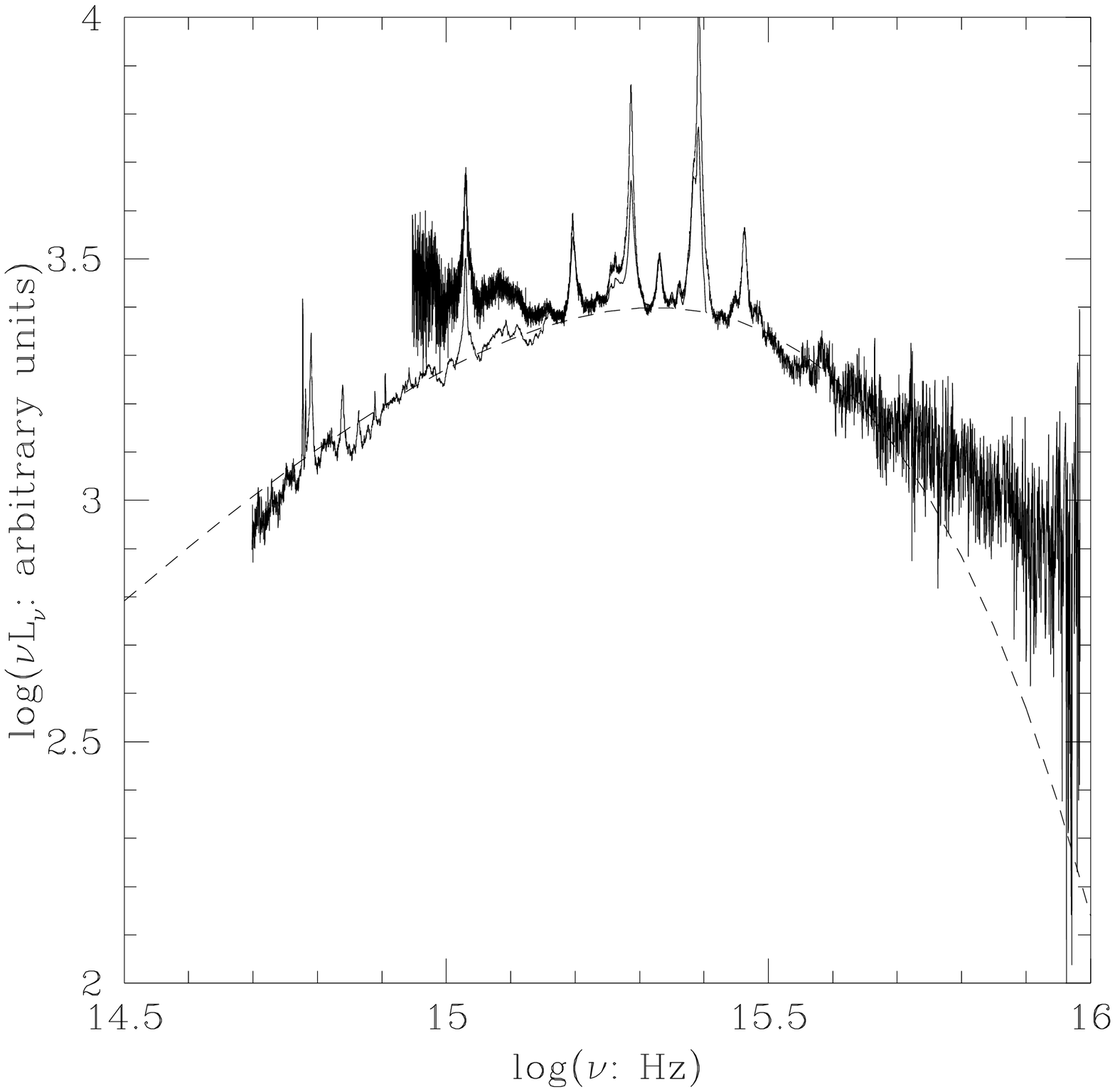}

\caption{Match between a zeroth-order predicted disk spectrum and two
composites of quasar spectra.  The noisy solid curves are the composites
assembled by Francis \etal\ (1991) (from the LBQS sample) and Zheng
\etal (1997) (from the FOS archive); the dashed curve has the form
$\nu^{4/3}\exp{(-h\nu/kT)}$ with $T = 7 \times 10^4$~K.}

\end{figure}

  By contrast, the more detailed spectral predictions must be tuned to
match the low frequency spectral shape, and nearly always show Lyman edge features, whether in emission or absorption.  While it is possible to smooth
out the edge with relativistic effects, it seems
unlikely that the feature can disappear altogether in composite spectra
unless polar views are somehow forbidden.

   It is possible that Comptonization can solve two problems---the strength
of the EUV continuum and the absence of Lyman edges, but it faces tough
prerequisites.  As Laor \etal\ (1997) have shown, the slope of the EUV spectrum
as seen at the highest frequencies observable in the ultraviolet is a good
predictor of the soft X-ray continuum, suggesting that this component
extends smoothly all the way across the unobservable EUV.  If this is
true (the interpolation in Figure 1 assumes this), a significant fraction
of the total luminosity of AGN is emitted in the EUV.  To make this
segment of the spectrum by Comptonization therefore entails putting the same
fraction of the total dissipation into the Comptonizing layer.  At present
there is no known physical mechanism that will do so.

   Thus, the interim conclusion regarding how well our models of accretion
disks fare vis-a-vis observations is one that is frustrating in several
respects.  On
the one hand, there are several major conceptual issues we do not know how
to settle that introduce large uncertainties into our predictions; on
the other hand, none of the guesses made to date on how to resolve these
questions leads to predictions that match observations.  It seems quite
likely that we are still missing a large piece of the puzzle.

\begin{acknowledgments}

     I thank Eric Agol and Mark Sincell for many enlightening conversations,
and for supplying figures, sometimes from unpublished work.   My
research on accretion disks is partially supported by NASA Grant
NAG5-3929 and by NSF Grant AST-9616922.

\end{acknowledgments}

\end{document}